\documentclass[aps,prx,twocolumn,superscriptaddress,citeautoscript]{revtex4-1}

\usepackage{amsmath,amssymb}
\usepackage{bm}
\usepackage{graphicx}
\usepackage{epstopdf}
\usepackage{latexsym}
\usepackage{subfigure}
\usepackage{color}
\usepackage{dsfont} 
\usepackage{wasysym} 

\newcommand{\be}{\begin{equation}}
\newcommand{\ee}{\end{equation}}
\newcommand{\bea}{\begin{eqnarray}}
\newcommand{\eea}{\end{eqnarray}}

\newcommand{\yto}{{Yb$_2$Ti$_2$O$_7$}}

\begin{document}

\title{Coulombic Quantum Liquids in Spin-1/2 Pyrochlores}

\author{Lucile Savary}
\affiliation{Ecole Normale Sup\'{e}rieure de Lyon,\, 46,\ all\'{e}e d'Italie, 69364 Lyon Cedex 07, France}
\affiliation{Department of Physics, University of California, Santa Barbara, CA 93106-9530, U.S.A.}
\author{Leon Balents}
\affiliation{Kavli Institute for Theoretical Physics, University of
  California, Santa Barbara, CA, 93106-4030, U.S.A.}

\date{\today}
\begin{abstract}
  We develop a non-perturbative ``gauge Mean Field Theory'' (gMFT)
  method to study a general effective spin-$1/2$ model for magnetism in
  rare earth pyrochlores.  gMFT is based on a novel exact slave-particle
  formulation, and matches both the perturbative regime near the
  classical spin ice limit and the semiclassical approximation far from
  it.  We show that the full phase diagram contains two exotic phases: a
  quantum spin liquid and a coulombic ferromagnet, both of which support
  deconfined spinon excitations and emergent quantum
  electrodynamics. Phenomenological properties of these phases are
  discussed.  %
\end{abstract}

\maketitle

Amongst the celebrated exotic phases of matter, of particular recent
interest are the Quantum Spin Liquids (QSLs) \cite{balents2010spin}.  Behind {\sl seemingly}
innocuous defining properties --strong spin correlations, the absence
of static magnetic moments, and unbroken crystalline symmetry--, QSLs
display the consequences of {\sl extreme} quantum entanglement.  These
include emergent gauge fields and fractional excitations, which take
these states beyond the usual ``mean field'' paradigm of phases of
matter.  Not only are these phases challenging to predict and describe,
they have also proven very hard to find in the laboratory, 
rendering their search and discovery even more tantalizing.

A consensual place to look for QSLs is among frustrated
magnets \cite{balents2010spin}.  Frustration allows the spins to
avoid  phases where they are either ordered or frozen,
with relatively small fluctuations and correlations between them.
Recent experiments have given compelling evidence of a QSL state in
certain two-dimensional organic materials \cite{kanoda11:_mott_physic_organ_conduc_trian_lattic}, but
both microscopic and fully consistent phenomenological theories are
lacking.  By contrast, {\sl classical} spin liquids have been
conclusively seen and microscopically understood in the spin ice
pyrochlores \cite{gingras11:_spin_ice}.  This raises
the possibility, suggested experimentally \cite{jointpaper} and theoretically \cite{tb2ti2o7-gingras-canals}, of QSLs
in those rare earth pyrochlores in which spins are non-classical,
supported by recent results on \yto\ \cite{jointpaper}.  However, for
any material, only detailed, quantitative theory predicting the {\sl
  type(s)} and properties of QSLs that appear {\sl and} matching
experiments can take the physics to the next level.

\begin{figure}[h]
\begin{center}
\includegraphics[width=3.3in]{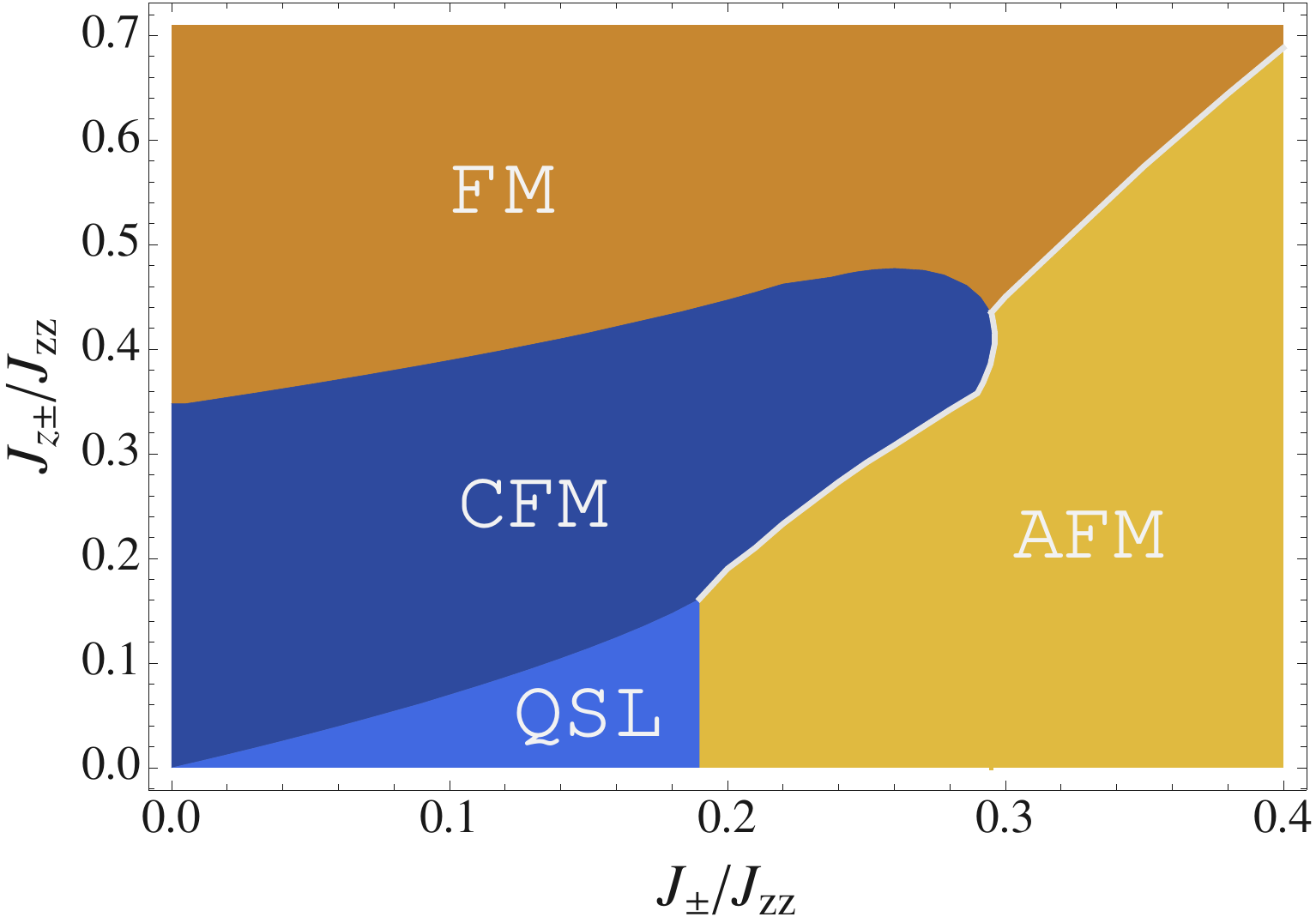}
\caption{Gauge mean field phase diagram obtained for $J_{\pm\pm}=0$ and
  $J_{zz}>0$.  ``QSL'', ``CFM'', ``FM'' and ``AFM'' denote the $U(1)$
  Quantum Spin Liquid, Coulomb Ferromagnet, standard ferromagnet, and
  standard antiferromagnet, respectively.  Phase boundaries with/without
  white lines indicate continuous/discontinuous transitions in gMFT.  Note that the diagram is symmetric in $J_{z\pm}\rightarrow-J_{z\pm}$. }
\label{fig:MF-phasediag}
\end{center}
\end{figure}

We take up this challenge here for quantum rare earth pyrochlores.  Our
analysis confirms that a ``$U(1)$'' QSL phase exists in the phase
diagram (Fig.~\ref{fig:MF-phasediag}) of a spectrum of real materials,
and is furthermore supplemented by another exotic phase, a Coulombic
ferromagnet, which contains spinons, but displays non-zero
magnetization.  We also study the confinement transitions out of these
Coulomb phases, which are analogous to ``Higgs''
transitions \cite{fradkin-shenker}.  Finally, we discuss experimental
signatures of the $U(1)$ QSL, and of the $U(1)$ Coulomb ferromagnet.

The most general nearest-neighbor
symmetry-allowed exchange Hamiltonian for spin-$1/2$ spins (real or
effective) on the pyrochlore lattice is
\begin{eqnarray}
  \label{eq:spinham}
  H & = & \sum_{\langle ij\rangle} \Big[ J_{zz} \mathsf{S}_i^z \mathsf{S}_j^z - J_{\pm}
  (\mathsf{S}_i^+ \mathsf{S}_j^- + \mathsf{S}_i^- \mathsf{S}_j^+) \nonumber \\
  && +\, J_{\pm\pm} \left[\gamma_{ij} \mathsf{S}_i^+ \mathsf{S}_j^+ + \gamma_{ij}^*
    \mathsf{S}_i^-\mathsf{S}_j^-\right] \nonumber \\
&& +\, J_{z\pm}\left[ \mathsf{S}_i^z (\zeta_{ij} \mathsf{S}_j^+ + \zeta^*_{ij} \mathsf{S}_j^-) +
  {i\leftrightarrow j}\right]\Big],
\end{eqnarray}
where $\gamma$ is a $4\times4$ complex unimodular matrix, and $\zeta=-\gamma^*$. The explicit expression of $\gamma$ and of the local bases whose
components are used in Eq.~\eqref{eq:spinham} are given in the
Supplementary Material.  The first term (we assume in this paper $J_{zz}>0$),
taken alone, gives the highly frustrated classical nearest-neighbor spin
ice model, which exhibits an extensive ground state degeneracy of
``two-in-two-out'' states. 

In fact, this model has been studied theoretically in the special case
$J_{z\pm}=J_{\pm\pm}=0$, where it reduces to an ``XXZ'' model with
global XY spin-rotation symmetry \cite{hermele}.  There, it was shown that for $J_\pm
\ll J_{zz}$, it is perturbatively equivalent, order by order, to a
lattice $U(1)$ gauge theory, with gauge fields that describe the spin
configurations {\sl constrained} to the spin ice manifold of ground
states.  This gauge theory was furthermore argued
to exhibit a so-called ``Coulomb phase'', which corresponds to a $U(1)$
QSL phase.  Subsequent numerical
simulations \cite{banerjee,shannon-sikora-2011} verified this prediction.
This Coulombic QSL is not only magnetically disordered, but also
supports several exotic excitations: spinons (called magnetic monopoles
in the spin ice literature), dual ``electric monopoles'', and an
emergent photon.  This understanding, however, was limited to the
perturbative regime $J_\pm \ll J_{zz}$ and considered only the XXZ case.
Here we develop a {\sl non-perturbative} method to analyze the full
Hamiltonian in Eq.~\eqref{eq:spinham}.  

Non-perturbative theories of QSLs based on ``slave particles'' have been
developed and used extensively in $SU(2)$ invariant $S=1/2$ Heisenberg
and Hubbard models \cite{wen2004quantum}.  Generally these approaches work by embedding the
Hilbert space on each site in some larger ``spinon'' one, with a microscopic gauge symmetry which acts to project back to
the physical space.  QSL phases are found when, in a mean field sense,
this microscopic gauge symmetry is incompletely broken in the ground
state.  Here, we follow the spirit but not the letter of these
approaches, by introducing redundant degrees of freedom not for each
spin but for each tetrahedron of the pyrochlore lattice.  This new slave
particle representation is, like the aforementioned standard ones,
formally exact, but additionally naturally describes the
Coulombic QSL found before in the perturbative analysis, when that limit
is taken.  It also has the added advantage that, unlike in standard
approaches, the gauge fields appear explicitly in the slave particle
Hamiltonian, rendering the analogy to lattice gauge theory more direct
and transparent. 

By dint of the theory developed in Refs. \onlinecite{hermele,
  banerjee,jointpaper}, we define our slave particles on the centers of
the ``up'' and ``down'' tetrahedra of the pyrochlore lattice, which
comprise two FCC sublattices (I/II, with $\eta_\mathbf{r}=\pm1$) of sites, denoted with
boldface characters $\mathbf{r}$, of a dual diamond lattice.  The sites
of the original pyrochlore lattice are bonds of the dual lattice.  The
perturbative analysis of Ref.~\onlinecite{hermele} identified the low
energy states of $H$ as the spin ice ones, supplemented by spinons
corresponding to defect tetrahedra.  As mentioned above, this inspires us to
enlarge the Hilbert space and define ``spinon'' slave
operators, which in turn can be seen as particles in a fluctuating
vacuum (the two-in-two-out manifold dear to the spin ice
  community).  We consider
$\mathcal{H}_{big}=\mathcal{H}_{spin}\otimes\mathcal{H}_{Q}$, where
$\mathcal{H}_{spin}=\bigotimes_N\mathcal{H}_{1/2}$ is the Hilbert space
of Eq.~\eqref{eq:spinham} and $\mathcal{H}_{Q}$ is the Hilbert space of
a field $Q_\mathbf{r}\in\mathbb{Z}$. $Q_\mathbf{r}$ is defined on all
the sites of the dual diamond lattice and, at this stage, is free and
unphysical.  We further define the real and compact operator
$\varphi_\mathbf{r}$ to be the canonically conjugate variable to
$Q_\mathbf{r}$, $[\varphi_\mathbf{r},Q_\mathbf{r}]=i$. In
$\mathcal{H}_{Q}$, the bosonic operators
$\Phi_\mathbf{r}^\dagger=e^{i\varphi_\mathbf{r}}$ and
$\Phi_\mathbf{r}=e^{-i\varphi_\mathbf{r}}$ thus act as raising and
lowering operators, respectively, for $Q_\mathbf{r}$.  Note that, by
construction, $|\Phi_\mathbf{r}|=1$.  We now take the restriction of
$\mathcal{H}_{big}$ to the subspace $\mathcal{H}$, in which
\begin{equation}
  \label{eq:gauss}
  Q_\mathbf{r} = \eta_\mathbf{r}
    \sum_\mu {\sf
      s}^z_{\mathbf{r},\mathbf{r}+\eta_\mathbf{r}\mathbf{e}_\mu},
\end{equation}
where the $\mathbf{e}_\mu$'s are the four
nearest-neighbor vectors of an $\eta_\mathbf{r}=1$ (I) diamond sublattice site.  This
constraint can be viewed as analogous to Gauss' law, where now
$Q_\mathbf{r}$ counts the number of spinons. The restriction of
$Q_\mathbf{r}$,  $\Phi_\mathbf{r}$ and
$\Phi_\mathbf{r}^\dagger$ to $\mathcal{H}$ exactly reproduces all matrix
elements of the original $\mathcal{H}_{spin}$, with the replacements
\begin{equation}
  \label{eq:1}
  \mathsf{S}_{\mathbf{r},\mathbf{r}+\mathbf{e}_\mu}^+ =
\Phi_{\mathbf{r}}^\dagger\,
\mathsf{s}_{\mathbf{r},\mathbf{r}+\mathbf{e}_\mu}^+
\Phi_{\mathbf{r}+\mathbf{e}_\mu}, \qquad
\mathsf{S}_{\mathbf{r},\mathbf{r}+\mathbf{e}_\mu}^z =
\mathsf{s}_{\mathbf{r},\mathbf{r}+\mathbf{e}_\mu}^z.
\end{equation}
Here $\mathbf{r}\in\mbox{I}$, and $\mathsf{s}^\pm_{\mathbf{rr}'}, \mathsf{s}^z_{\mathbf{rr}'}$ act
within the $\mathcal{H}_{spin}$ subspace of $\mathcal{H}_{big}$.
Note especially that, by itself, $\mathsf{s}^\pm_{\mathbf{rr}'} \neq
\mathsf{S}^\pm_{\mathbf{rr}'}$ is not the physical spin, and does not
remain within $\mathcal{H}$.

In this paper we focus on the case where $J_{\pm\pm}=0$ (which
otherwise introduces additional complications to be dealt with in a separate
publication), and the Hamiltonian then becomes
\begin{widetext}
  \begin{eqnarray}
    \label{eq:ham-spinons}
    H & = & \sum_{\mathbf{r} \in {\rm I,II}} \frac{J_{zz}}{2} Q_\mathbf{r}^2 - J_\pm \left\{ \sum_{\mathbf{r}\in
       {\rm I}}  \sum_{\mu,\nu\neq\mu}  \Phi_{\mathbf{r}+\mathbf{e}_\mu}^\dagger
      \Phi_{\mathbf{r}+\mathbf{e}_\nu}^{\vphantom\dagger} {\sf
        s}^-_{\mathbf{r},\mathbf{r}+\mathbf{e}_\mu} {\sf
        s}^+_{\mathbf{r},\mathbf{r}+\mathbf{e}_\nu} + \sum_{\mathbf{r}\in
       {\rm II}}  \sum_{\mu,\nu\neq\mu} \Phi_{\mathbf{r}-\mathbf{e}_\mu}^\dagger
      \Phi_{\mathbf{r}-\mathbf{e}_\nu}^{\vphantom\dagger} {\sf s}^+_{\mathbf{r},\mathbf{r}-\mathbf{e}_\mu} {\sf s}^-_{\mathbf{r},\mathbf{r}-\mathbf{e}_\nu}
       \right\} \\
  && - J_{z\pm} \left\{ \sum_{r\in {\rm I}} \sum_{\mu,\nu\neq\mu} \left(
    \gamma^*_{\mu\nu} \Phi_\mathbf{r}^\dagger \Phi_{\mathbf{r}+\mathbf{e}_\nu}^{\vphantom\dagger}
    {\sf s}^z_{\mathbf{r},\mathbf{r}+\mathbf{e}_\mu} {\sf s}^+_{\mathbf{r},\mathbf{r}+\mathbf{e}_\nu} + {\rm h.c.}\right)
  + \sum_{\mathbf{r}\in {\rm II}} \sum_{\mu,\nu\neq\mu} \left(
    \gamma^*_{\mu\nu} \Phi_{\mathbf{r}-\mathbf{e}_\nu}^\dagger \Phi_{r}^{\vphantom\dagger}
    {\sf s}^z_{\mathbf{r},\mathbf{r}-\mathbf{e}_\mu} {\sf s}^+_{\mathbf{r},\mathbf{r}-\mathbf{e}_\nu} + {\rm h.c.}\right) \right\}+\mbox{const.}.
\nonumber  
\end{eqnarray}
\end{widetext}
The integer-valued constraint in Eq.~\eqref{eq:gauss} commutes with $H$ and thereby ensures that
Eq.~\eqref{eq:ham-spinons} is a $U(1)$ gauge theory.  Explicitly, it is
invariant under the transformations
\begin{equation}
\label{eq:gauge-sym}
\begin{cases}
\Phi_\mathbf{r}\rightarrow\Phi_\mathbf{r}\,e^{-i\chi_\mathbf{r}}\\
{\sf s}^\pm_{\mathbf{rr}'}\rightarrow {\sf s}_{\mathbf{rr}'}^\pm e^{\pm i(\chi_{\mathbf{r}'}-\chi_{\mathbf{r}\vphantom{'}})}
\end{cases},
\end{equation}
with arbitrary $\chi_\mathbf{r}$.  This invariance, and the Gauss' law
in Eq.~\eqref{eq:gauss} can be made formally identical to that in
lattice electrodynamics by writing
$\mathsf{s}^z_{\mathbf{rr}'}=E_{\mathbf{rr}'}$ and
$\mathsf{s}^\pm_{\mathbf{rr}'}= e^{\pm iA_{\mathbf{rr}'}}$, where $E$
and $A$ are lattice electric and magnetic fields \cite{hermele}.  This
clarifies that $\mathsf{s}^\pm_{\mathbf{rr}'}$ is to be regarded as an
element of the $U(1)$ gauge group.  However, the notation is
unnecessary and we use it only when conceptually valuable.

Eq.~\eqref{eq:ham-spinons} can be viewed as spinons hopping in the
background of fluctuating gauge fields, and thereby lends itself to the
application of standard mean field theory methods for lattice gauge
models \cite{PhysRevD.10.2445}, which we call gauge Mean Field Theory (gMFT).  Upon performing gMFT, we
will get a Hamiltonian for spinons hopping in a fixed background.
Specifically, we perform the replacement:
\begin{eqnarray}
\label{eq:MFTreplace}
&&\Phi^\dagger\Phi\,\mathsf{s}\,\mathsf{s}\rightarrow\\
&&\qquad\Phi^\dagger\Phi\langle\mathsf{s}\rangle\langle\mathsf{s}\rangle+\langle\Phi^\dagger\Phi\rangle\mathsf{s}\langle\mathsf{s}\rangle+\langle\Phi^\dagger\Phi\rangle\langle\mathsf{s}\rangle\mathsf{s}-2\langle\Phi^\dagger\Phi\rangle\langle\mathsf{s}\rangle\langle\mathsf{s}\rangle,\nonumber
\end{eqnarray}
and thereby split the Hamiltonian in a spinon part $H_\Phi^{{\rm MF}}$,
and a gauge part $H_g^{{\rm MF}}$.  Note that unlike conventional
Curie-Weiss mean field theory, which entirely neglects any quantum
entanglement, gMFT, while suppressing some fluctuations, still allows
high correlations and entanglement.

The gMFT order parameters
are closely analogous to those in $U(1)$ Higgs
theory \cite{PhysRevD.10.2445,PhysRevD.28.360}.  A non-zero expectation
value $\langle \mathsf{s}^\pm\rangle \neq 0$ implies the phase of
$\mathsf{s}^\pm$ is relatively well-defined, i.e. there are small
fluctuations of the vector potential $A$.  The converse case, $\langle
\mathsf{s}^\pm\rangle =0$ would indicate confinement, but does not occur
here.  A non-zero scalar expectation value, $\langle \Phi\rangle \neq
0$, analogous to a Higgs phase, indicates spinon condensation and
generation of a mass for the gauge field, and a conventional, non-exotic
state.  Combined with $\langle \mathsf{s}^\pm\rangle \neq 0$, it also
implies ``XY'' magnetic order.  Conversely, $\langle \Phi\rangle =0$
indicates the spinons have a gap, and is characteristic of the Coulomb
phase.  The remaining gMFT order parameter, $\mathsf{s}^z$, is gauge
invariant, and thus indicates only the presence ($\langle
\mathsf{s}^z\rangle\neq 0$) or absence ($\langle \mathsf{s}^z\rangle=
0$) of ``Ising'' magnetic order, i.e. time-reversal symmetry breaking.
Combining this together, the phases in gMFT are summarized in
Table~\ref{tab:phase-class}.
\begin{table}[htdp]
  \caption{Order parameters and phases in gMFT. }
\begin{center}
\begin{tabular}{|c|c|c||c|}
\hline \quad$\langle\Phi\rangle\quad$ & \quad$\langle \mathsf{s}^z \rangle\quad$ & \quad$\langle \mathsf{s}^\pm\rangle\quad$ & \quad phase$\quad$  \\ \hline 
$0$ & $0$ & $\neq0$ & QSL \\ \hline
$0$ & $\neq0$ & $\neq0$ & CFM \\ \hline
$\neq0$ & $\neq0$ & $\neq0$ & FM \\ \hline
$\neq0$ & $0$ & $\neq0$ & AFM \\ \hline
\end{tabular}
\end{center}
\label{tab:phase-class}
\end{table}%
We emphasize that despite the fact that $\langle \Phi\rangle$ does not
appear explicitly in the decoupling in Eq.~\eqref{eq:MFTreplace}, the
gMFT does generally allow for Higgs phases where $\Phi$ is indeed
condensed.  As we will show below, the Higgs phase appears in a manner
similar to Bose-Einstein condensation in an ideal Bose gas. 

We now use the following Ansatz, valid when $J_\pm>0$ (which we assume hereafter),
\begin{equation}
\langle\mathsf{s}_\mu^z\rangle=\frac{1}{2}\sin\theta\,\varepsilon_\mu,\qquad\langle\mathsf{s}^-_\mu\rangle=\frac{1}{2}\cos\theta,
\label{eq:ansatz}
\end{equation}
where $\mu=0,..,3$ and $\varepsilon=(1,1,-1,-1)$, which assumes
translational invariance and fully polarized ``spins''
$\vec{\mathsf{s}}$, in accord with Eq.~\eqref{eq:MFTreplace}, and is
compatible with FM polarization along the (global) $x$ axis
($\langle\mathsf{s}^+_\mu\rangle=\langle\mathsf{s}^-_\mu\rangle$).  Note
that Eq.~\eqref{eq:ansatz} shows that the gMFT allows fluctuations of
$E$ and $A$, so long as $\theta\neq\pi/2$ and $\theta\neq0$,
respectively. Defining the dot product, and through it the vector
notation,
$\vec{\mathsf{u}}\cdot\vec{\mathsf{v}}=\mathsf{u}^z\mathsf{v}^z+\frac{1}{2}\left(\mathsf{u}^+\mathsf{v}^-+\mathsf{u}^-\mathsf{v}^+\right)$,
we find 
\begin{equation}
H^{\rm MF}_{\sf s} =- \sum_{\mathbf{r}\in {\rm I}} \sum_\mu \vec{{\sf h}}_{{\rm eff},\mu}(\mathbf{r})\cdot \vec{{\sf s}}_{\mathbf{r},\mathbf{r}+\mathbf{e}_\mu},
\end{equation}
where ${\sf h}_{{\rm eff},\mu}^z = 4\, \varepsilon_\mu J_{z\pm}
I_1\cos\theta$ and ${\sf h}_{{\rm eff},\mu}^- = 4 J_{z\pm}
I_1\sin\theta +2J_\pm I_2\cos\theta$, and we have defined
$I_1=\varepsilon_\mu\langle\Phi_\mathbf{r}^\dagger\Phi^{\vphantom\dagger}_{\mathbf{r}+\mathbf{e}_\mu}\rangle$
(no summation implied) and
$I_2=\sum_{\nu\neq\mu}\langle\Phi_\mathbf{r}^\dagger\Phi^{\vphantom\dagger}_{\mathbf{r}+\mathbf{e}_\mu-\mathbf{e}_\nu}\rangle$
($\mu$ is fixed).  These quantities turn out to be independent of the
diamond bond $\mu$.  To treat the spinons, we relax the
$|\Phi_\mathbf{r}|=1$ constraint to a global one by introducing a
Lagrange multiplier $\lambda$ via the term
$\lambda\sum_\mathbf{r}\left(|\Phi_\mathbf{r}|^2-1\right)$ in a path
integral formulation, with free integration over $\Phi$ and $\Phi^*$.
The spinon Lagrangian is
\begin{equation}
\mathcal{L}^{{\rm MF}}_\Phi=\frac{1}{N_{{\rm u.c.}}}\sum_{\mathbf{k}}\int_{\omega_n}\Phi_{\mathbf{k},\omega_n}^*\cdot G^{-1}_{\mathbf{k},\omega_n}\cdot\Phi^{\vphantom*}_{\mathbf{k},\omega_n},
\end{equation}
where $N_{{\rm u.c.}}$ is the number of unit cells, $\left[G_{\mathbf{k},\omega_n}\right]_{ab}=\langle\Phi_b^*\Phi_a\rangle$, and we find the equal time Green's function to be
\begin{equation}
\label{eq:green-fn}
G_{\mathbf{k},\tau=0}=\frac{1}{2}\sqrt{\frac{J_{zz}}{2}}
\begin{pmatrix}
Z_\mathbf{k}^+ & -\frac{M_\mathbf{k}}{|M_\mathbf{k}|}Z^-_\mathbf{k}\\
-\frac{M_\mathbf{k}^*}{|M_\mathbf{k}|}Z^-_\mathbf{k} & Z_\mathbf{k}^+
\end{pmatrix},
\end{equation}
where $M_\mathbf{k}=\sum_\mu\varepsilon_\mu e^{i\mathbf{k}\cdot
  \mathbf{e}_\mu}$,
$Z_\mathbf{k}^\pm\left(\theta,\lambda\right)=\frac{1}{z_\mathbf{k}^+}\pm\frac{1}{z_\mathbf{k}^-}$,
$z_\mathbf{k}^\pm\left(\theta,\lambda\right)=\sqrt{\lambda-\ell_\mathbf{k}^\pm(\theta)}$,
$\ell^\pm_\mathbf{k}(\theta)=\frac{1}{2}J_\pm\cos^2\theta
L_\mathbf{k}\mp\left|\frac{1}{2}J_{z\pm}\sin2\theta
  M_\mathbf{k}\right|$,
$L_\mathbf{k}=\sum_{\mu,\nu<\mu}\cos\left[\mathbf{k}\cdot(\mathbf{e}_\mu-\mathbf{e}_\nu)\right]$.
A couple of remarks are in order: (i)
$\lambda>\max_\mathbf{k}\ell_\mathbf{k}^-$ (ii) the spinon dispersion
relations are
$\omega^\pm_\mathbf{k}(\theta,\lambda)=\sqrt{2J_{zz}}\,z^\pm_\mathbf{k}(\theta,\lambda)$.

The gMFT consistency conditions on $\theta$ and $\lambda$ (for fixed $J_\pm,J_{z\pm}$) arise from
requiring Eqs.~\eqref{eq:ansatz}  and $\langle \mathsf{s}^\kappa_\mu \rangle
= \mathsf{h}^\kappa_\mu/(2|\vec{\mathsf{h}}_\mu|)$, and from the normalization
condition on $|\Phi|^2$, and can be written
\begin{equation}
\begin{cases}\tan\theta=\dfrac{2J_{z\pm}I_1(\theta,\lambda)}{2J_{z\pm}I_1(\theta,\lambda)\tan\theta+J_\pm I_2(\theta,\lambda)}\\
I_3\left(\theta,\lambda\right)=1\end{cases},
\label{eq:consistency}
\end{equation}
where
$I_3\left(\theta,\lambda\right)=\langle\Phi_\mathbf{r}^\dagger\Phi_\mathbf{r}^{\vphantom{\dagger}}\rangle$.
The explicit expressions for the $I_i=N_{{\rm
      u.c.}}^{-1}\sum_\mathbf{k}\mathcal{I}^i_\mathbf{k}$, needed to solve
Eqs.~\eqref{eq:consistency}, are readily derived from
Eq.~\eqref{eq:green-fn}, and are given in the Supplementary Material,
Eq.~\eqref{eq:integrals}.  Since Eqs.~\eqref{eq:consistency} may allow
several distinct solutions, we must choose the solution of
Eq.~\eqref{eq:consistency} with the lowest energy.  In the mean field
approximation, the ground state energy can be calculated by taking the
expectation value of the Hamiltonian.  We find, per unit cell,
$\epsilon_{\rm GS}=\epsilon_{av}+\epsilon_{kin}$, with
\begin{eqnarray}
\label{eq:energy-av}
\epsilon_{av}&=&-2I_2(\theta,\lambda)\cos^2\theta
J_\pm-4I_1(\theta,\lambda)\sin2\theta\, J_{z\pm}\\
\epsilon_{kin}&=&\frac{1}{2}\int_\mathbf{k}\left[\omega^+_\mathbf{k}(\theta,\lambda)+\omega^-_\mathbf{k}(\theta,\lambda)\right],
\label{eq:energy-kin}
\end{eqnarray}
where $\omega^\pm=\sqrt{2J_{zz}}\,z^\pm$. Here $\epsilon_{kin}$ measures the ``kinetic'' energy associated with the spinon modes, while $\epsilon_{av}$ represents the ``background'' energy in which the latter evolve.

We now discuss how the different phases are obtained from the solutions of the
gMFT equations.  Condensed and uncondensed phases are distinguished by
the value of $\lambda$.  As in the theory of superfluidity,
condensation is synonymous with off-diagonal long-range order,
i.e. $\langle \Phi_\mathbf{r}\rangle^* \langle
\Phi_{\mathbf{r}'}\rangle \equiv
\lim_{|\mathbf{r}-\mathbf{r}'|\rightarrow\infty}
\langle\Phi_\mathbf{r}^\dagger\Phi_{\mathbf{r}'}^{\vphantom{\dagger}}\rangle
\neq 0$.  This expectation value
$\langle\Phi_{\mathbf{r}'}^{\vphantom{\dagger}}\Phi_\mathbf{r}^\dagger\rangle=N_{{\rm
      u.c.}}^{-1}\sum_\mathbf{k}G_\mathbf{k}e^{i\mathbf{k}\cdot(\mathbf{r}'-\mathbf{r})}$ is non-zero in the long-distance limit if and only if the usual
conversion of the sum to an integral {\sl fails}, i.e. if there exists
one $\mathbf{k}_0$ such that $G_{\mathbf{k}_0}=O(N_{{\rm u.c.}})$.
Like the chemical potential in Bose-Einstein condensation, $\lambda$
in a Higgs phase differs from its minimum allowed value by a
sub-extensive part. $\lambda=\lambda_{{\rm
    min}}(\theta)+\frac{\delta^2}{N_{{\rm u.c.}}^2}$ therefore defines
condensation, where $\delta$ is of order $O(1)$, and $\lambda_{{\rm
    min}}(\theta)=\max_\mathbf{k}\ell_\mathbf{k}^-(\theta)$.  If
instead $\lambda-\lambda_{{\rm min}}(\theta)$ is $O(1)$, one has a
phase with deconfined gapped spinons and a Coulombic gauge structure.
As we already mentioned above, these classes of phases can be further
subdivided into ``polarized'' (i.e. with magnetization along the {\sl local}
$z$ axis) and ``unpolarized'' when $\theta\neq0$ and $\theta=0$,
respectively.

The phase diagram resulting from the gMFT solution (see Supplementary
material) is shown in Fig.~\ref{fig:MF-phasediag}.  It contains two
``exotic'' phases in which spinons are deconfined and uncondensed,
indicated as QSL and CFM.  The QSL state, with $\theta=0$, is completely
absent magnetic order, and is the phase studied in
Refs.~\onlinecite{hermele,banerjee}.  Its low energy physics mimics
quantum electrodynamics, and thereby contains a photonic excitation
(gapless and linear near the origin) and gapped fractional monopole
excitations (spinon and ``electric'' monopole) that interact via Coulomb
interactions.  In the present formalism, the photon is only obtained
once quadratic fluctuations around the gMFT solution are considered, but
is a universal feature of the exotic phases.  The CFM, or ``Coulombic
Ferromagnet'' phase, is a new phase of matter that can be seen as a
polarized version of the $U(1)$ QSL.  Despite being magnetic, its {\sl
  elementary} magnetic {\sl excitations} are spinons rather than spin
waves, and it also supports a gapless photon mode.  Indeed, in gMFT the
transition from the QSL to CFM is second order, and consists of a
continuous rise of magnetization from zero.  For larger
$J_{z\pm},J_\pm$, one obtains Higgs phases, which are conventional
states of matter without exotic excitations and are continuously
connected to the usual magnetically ordered states described by
Curie-Weiss MFT. Interestingly, we find the exotic CFM state is
considerably more stable than the ``pure'' QSL, occupying a much more
substantial portion of the phase diagram.  

How do we recognize a Coulomb phase in experiment?  A generic sign of
fractionalization is a two-particle continuum in inelastic neutron
scattering, two spinons being excited by one neutron
\cite{ross2009,jointpaper}.  In addition the photon can be detected
directly by inelastic neutron scattering, as a linearly dispersing
transverse mode.  It is, however, more challenging to observe than the
usual acoustic spin wave, because its scattering intensity becomes
small ($\propto\omega$) at low energy (see Supp. Mat.), in contrast to the spin wave
for which the intensity diverges ($\sim 1/\omega$) in the same limit. 
Interestingly, the pinch points in the static structure factor present
for classical spin ice are absent for the quantum Coulomb
phase \cite{shannon-sikora-2011}, so this is not a useful
measurement. Perhaps the most striking signature of the Coulomb phase
is likely to be thermodynamic.  Like the phonons, the photons
contribute as $B\, T^3$ to the specific heat at low temperatures, but
their speed is $v_{photon}\sim J\ll c$, the speed of sound.  Crudely
estimating $J\sim 2$~K appropriate for \yto, we obtain a coefficient
$B_{photon} \approx 10^3$mJ/mole-K$^4$, approximately 1000 times
larger (!) than the phonon contribution $B_{phonon} \approx 0.5$
mJ/mole-K$^4$ measured for the isostructural material
Y$_2$Ti$_2$O$_7$ \cite{PhysRevB.79.224111}. 

With a phase diagram and a new phase of matter in hand, we take heart at
discovering yet more new exciting facts in the pyrochore lattice.
Future studies should address the more frustrated case $J_\pm<0$, phase
transitions in applied field, and the influence of defects.
 
We thank Peter Holdsworth, Bruce Gaulin and Kate Ross for discussions.
This work was supported by the DOE through BES grant DE-FG02-08ER46524.

\bibliography{pyro_theory}

\appendix

\bigskip\bigskip




\begin{center}
{\bf SUPPLEMENTARY MATERIAL}
\end{center}

\section{Geometry}

\subsection{Local bases}

The local cubic bases in which the Hamiltonian Eq.~\eqref{eq:spinham} is expressed are the following $(\mathbf{\hat{a}}_i,\mathbf{\hat{b}}_i,\mathbf{\hat{e}}_i)$ bases 
\begin{equation}
\left\{\begin{array}{l}
\mathbf{\hat{e}}_0=(1,1,1)/\sqrt{3}\\
\mathbf{\hat{e}}_1=(1,-1,-1)/\sqrt{3}\\
\mathbf{\hat{e}}_2=(-1,1,-1)/\sqrt{3}\\
\mathbf{\hat{e}}_3=(-1,-1,1)/\sqrt{3},
\end{array}\right.,
\quad
\left\{\begin{array}{l}
\mathbf{\hat{a}}_0=(-2,1,1)/\sqrt{6}\\
\mathbf{\hat{a}}_1=(-2,-1,-1)/\sqrt{6}\\
\mathbf{\hat{a}}_2=(2,1,-1)/\sqrt{6}\\
\mathbf{\hat{a}}_3=(2,-1,1)/\sqrt{6}
\end{array}\right.,
\end{equation}
$\mathbf{\hat{b}}_i=\mathbf{\hat{e}}_i\times\mathbf{\hat{a}}_i$, such that spin $\mathbf{S}_i$ on sublattice $i$ is $\mathbf{S}_i=\mathsf{S}^+_i(\mathbf{\hat{a}}_i-i\mathbf{\hat{b}}_i)/2+\mathsf{S}^-_i(\mathbf{\hat{a}}_i+i\mathbf{\hat{b}}_i)/2+\mathsf{S}^z_i\mathbf{\hat{e}}_i$.

The $4\times4$ matrix $\gamma$ introduced in Eq.~\eqref{eq:spinham} is 
\begin{equation}
\gamma=\begin{pmatrix}
0 & 1 & w & w^2\\
1 & 0 & w^2 & w\\
w & w^2 & 0 & 1\\
w^2 & w & 1 & 0
\end{pmatrix},
\end{equation}
where $w=e^{2\pi i/3}$ is a third root of unity.

\subsection{Lattice vectors}

The four nearest-neighbor vectors of a I-sublattice diamond site (sublattice I corresponds to ``up'' tetrahedra) are $\mathbf{e}_\mu=\frac{a\sqrt{3}}{4}\mathbf{\hat{e}}_\mu$, where $a$ is the usual FCC lattice spacing. The four pyrochlore sites of the ``up'' tetrahedron centered at the origin are located at $\mathbf{e}_\mu/2$, $\mu=0,..,3$. 

The FCC primitive lattice vectors are $\mathbf{A}_i=\mathbf{e}_0-\mathbf{e}_i$, $i=1,..,3$, while the reciprocal lattice basis vectors are defined as usual by $\mathbf{B}_1=2\pi\frac{\mathbf{A}_2\times\mathbf{A}_3}{v_{{\rm u.c.}}}$ and its cyclic permutations, where $v_{{\rm u.c.}}=\mathbf{A}_1\cdot(\mathbf{A}_2\times\mathbf{A}_3)$ is the volume of the (real space) unit cell.  If the $q_i$'s are defined as 
\begin{equation}
\label{eq:qi}
\mathbf{k}=\sum_{i=1}^3 q_i\,\mathbf{B}_i,
\end{equation} 
the first Brillouin zone can be considered the ``cube'' with unit sides described by $-1/2<q_i<1/2$ (note that the $q_i$'s are dimensionless).


\section{Brillouin Zone Sums}

In the main text, we defined
\begin{eqnarray}
I_1&=&\varepsilon_\mu\langle\Phi_\mathbf{r}^\dagger\Phi_{\mathbf{r}+\mathbf{e}_\mu}\rangle\quad\mbox{(no summation implied)}, \\
I_2&=&\sum_{\nu\neq\mu}\langle\Phi_\mathbf{r}^\dagger\Phi_{\mathbf{r}+\mathbf{e}_\mu-\mathbf{e}_\nu}\rangle\quad\mbox{($\mu$ is fixed)},\\
I_3&=&\langle\Phi_\mathbf{r}^\dagger\Phi_\mathbf{r}^{\vphantom{\dagger}}\rangle.
\end{eqnarray}
For brevity, we define the $\mathcal{I}^i_\mathbf{k}$'s through
\begin{equation}
\label{eq:formal-sums}
I_i=\frac{1}{N_{{\rm u.c.}}}\sum_{\mathbf{k}\in{\rm BZ}}\mathcal{I}^i_\mathbf{k}=\frac{1}{N_{{\rm u.c.}}}\sum_{\mathbf{q}\in{\rm cube}}\mathcal{I}^i_\mathbf{q}
\end{equation}
($\mathbf{q}$ is defined in Eq.~\eqref{eq:qi}), whose explicit expressions are
\begin{eqnarray}
&&\mathcal{I}^1_\mathbf{q}\left(\theta,\lambda\right)=\nonumber\\
&&\quad\frac{1}{2}\sqrt{\frac{J_{zz}}{2}}\frac{Z^-_\mathbf{q}}{|M_\mathbf{q}|}\left(-1-\cos2\pi q_1+\cos2\pi q_2+\cos2\pi q_3\right)\nonumber,\\
&&\mathcal{I}^2_\mathbf{q}\left(\theta,\lambda\right)=\frac{1}{2}\sqrt{\frac{J_{zz}}{2}}Z^+_\mathbf{q}\left(\cos2\pi q_1+\cos2\pi q_2+\cos2\pi q_3\right),\nonumber\\
&&\mathcal{I}^3_\mathbf{q}\left(\theta,\lambda\right)=\frac{1}{2}\sqrt{\frac{J_{zz}}{2}}Z^+_\mathbf{q},
\label{eq:integrals}
\end{eqnarray}
where we used the Fourier transform convention $\Phi_\mathbf{r}=\frac{1}{N_{{\rm u.c.}}}\sum_\mathbf{k}\Phi_\mathbf{k}\,e^{i\mathbf{k}\cdot\mathbf{r}}$, and in turn 
\begin{equation}
\langle\Phi^\dagger_\mathbf{r}\Phi_{\mathbf{r}'}\rangle=\frac{1}{N_{u.c.}}\sum_\mathbf{k}[G_\mathbf{k}]_{21}e^{i\mathbf{k}\cdot(\mathbf{r}'-\mathbf{r})}\quad\mbox{for}\; \mathbf{r}\in\mbox{I},\mathbf{r}'\in\mbox{II},
\end{equation}
with $G_{\mathbf{k}}$ defined in Eq.~\eqref{eq:green-fn}.

These Brillouin zone sums need to be evaluated.  While in the Coulomb phases, they can be directly identified with the corresponding integrals, the situation is more complex within the condensed phases.  We address this issue below.

\section{Brillouin zone sums in the condensed phases}

The condensed region occurs if $I_3(\lambda_{{\rm min}})<1$.  In this
case, we must allow for the subextensive part of $\lambda$.  Let
\begin{equation}
  \label{eq:59}
  \lambda= \lambda_{\rm min} + \frac{\delta^2}{N_{u.c.}^2},
\end{equation}
where $N_{u.c.}$ is the
number of unit cells and $\delta>0$.  Then the sums Eq.~\eqref{eq:formal-sums} cannot merely be replaced by the corresponding Brillouin zone integrals, but must be written as follows:
\begin{equation}
  \label{eq:57}
\frac{1}{N_{u.c.}} \sum_\mathbf{k} f(\mathbf{k}) = \frac{f(\mathbf{k}_{\rm min})}{N_{u.c.}} + \frac{1}{N_{u.c.}} \sum_{\mathbf{k}\neq\mathbf{k}_{\rm min}} f(\mathbf{k})
\end{equation}
and
\begin{eqnarray}
&&\frac{1}{N_{u.c.}} \sum_{\mathbf{k}\neq\mathbf{k}_{\rm min}} f(\mathbf{k}) = \frac{V_{u.c.}}{V} \sum_{\mathbf{k}\neq\mathbf{k}_{\rm min}} f(\mathbf{k})\\
&&\quad = \frac{1}{V_{BZ}}\sum_{\mathbf{k}\neq\mathbf{k}_{\rm min}} \left(\frac{2\pi}{L}\right)^d f(\mathbf{k}) \rightarrow \int \! \frac{d^d k}{V_{BZ}} f(\mathbf{k})=\int_\mathbf{k} f(\mathbf{k}),\nonumber
\end{eqnarray}
where $V_{BZ}$ is the volume of the Brillouin zone.  In general, we denote this decomposion of an integral as $I= I_{\rm min}
+ I'$, where $I'$ is the continuous integral part evaluated at
$\lambda=\lambda_{\rm min}$, and $I_{\rm min}$ is the first term in the
right hand side of Eq.~\eqref{eq:57}.  In $I_{\rm min}$, we need only
retain the part which is non-vanishing as $N_{u.c.}\rightarrow \infty$.

The value $\lambda_{{\rm min}}$ is determined by the condition that $\min_\mathbf{q} z_\mathbf{q}^- =
0$, which corresponds to $\lambda_{{\rm min}} = \max_\mathbf{q}\ell_\mathbf{q}^-(\theta)=\max_\mathbf{q} (\frac{1}{2}\cos^2\theta J_\pm L_\mathbf{q}+ |\frac{1}{2}J_{z\pm}\sin2\theta M_\mathbf{q}|)$.  Here, this maximum $\max_\mathbf{q}\ell_\mathbf{q}^-$ always occurs for the (dimensionless) wavevector of the form $\mathbf{q}_{\rm min}=(q_1,q_2,q_3)=(0,q,q)$ ($\mathbf{q}_{\rm min}$ corresponds to $\mathbf{k}_0$ in the main text).  Defining
\begin{equation}
y=\cos 2\pi q,
\end{equation}
$\max_\mathbf{q}\ell_\mathbf{q}^-$ is realized for the wavevector $(0,q,q)$ corresponding to
\begin{equation}
  \label{eq:56}
  y_{{\rm min}} =
  \begin{cases}
    1 - \frac{1}{8}\left(\frac{J_2[J_{z\pm},\theta]}{J_1[J_\pm,\theta]}\right)^2 & \mbox{for } J_2[J_{z\pm},\theta]< 4 J_1[J_\pm,\theta]  \\
    -1 & \mbox{for }
    J_2[J_{z\pm},\theta]>4 J_1[J_\pm,\theta] 
  \end{cases},
\end{equation}
where we defined
\begin{eqnarray}
\label{eq:alpha}
J_1[J_{\pm},\theta]&=&\frac{1}{2}J_\pm\cos^2\theta, \\
J_2[J_{z\pm},\theta]&=&\frac{1}{2}J_{z\pm}\sin2\theta,
\label{eq:beta}
\end{eqnarray}
i.e. $\ell_\mathbf{k}^\pm(\theta)=J_1[J_\pm,\theta] L_\mathbf{k}\mp |J_2[J_{z\pm},\theta] M_\mathbf{k}|$, and assumed $J_2[J_{z\pm},\theta] >0$. Then we find
\begin{eqnarray}
  \label{eq:lambdamin}
 && \lambda_{{\rm min}} =\\
 &&\;\; \begin{cases}
    6 J_1[J_{\pm},\theta] + \frac{J_2[J_{z\pm},\theta]^2}{2J_1[J_{\pm},\theta]} & \mbox{for }
    J_2[J_{z\pm},\theta]< 4 J_1[J_{\pm},\theta] \\
    4 J_2[J_{z\pm},\theta]-2J_1[J_{\pm},\theta] & \mbox{for } J_2[J_{z\pm},\theta]> 4 J_1[J_{\pm},\theta]
  \end{cases}.\nonumber
\end{eqnarray}

There is one subtlety.  When $\theta=0$, $z^+_\mathbf{k}=z^-_\mathbf{k}$, so that
there will be a degeneracy of the minimum energy state.  This changes
the condensed contributions.  The two cases can actually be treated together, provided we formulate
everything in terms of a new variable $\rho$, the ``condensed density'', rather than $\delta$, where
\begin{equation}
\begin{cases}
\rho(\theta>0)=\frac{1}{2}\sqrt{\frac{J_{zz}}{2}} \frac{1}{\delta} \\
\rho(\theta=0) =\sqrt{\frac{J_{zz}}{2}} \frac{1}{\delta}
\end{cases},
\end{equation}
i.e. the condensed density is larger in the case $\theta=0$ by a factor of $2$.  Then, from Eqs.~\eqref{eq:consistency} and \eqref{eq:integrals}, we find
\begin{eqnarray}
  \label{eq:60}
  \rho &=& 1 - I'_3,\\
  \label{eq:62}
  I_1 &=& \frac{\rho}{\sqrt{2}} \sqrt{1-y_{\rm min}} + I'_1,\\
  I_2 &=& \rho(1+2y_{\rm min}) + I'_2.
  \label{eq:63}
\end{eqnarray}
When $\theta=0$, the minimum is reached at $\mathbf{q}_{\rm min}=\mathbf{0}$, i.e. $y_{\rm min}=1$. Eq.~\eqref{eq:62} then reduces to $I_1'(\theta=0)=0$ since $I_1(\theta=0)$ vanishes identically (which reflects the fact that the two FCC sublattices are decoupled), and Eq.~\eqref{eq:63} reduces to $I_2 = 3\rho+ I'_2$.

\section{Ground State Energy Calculation}


In some cases, we find multiple solutions of the mean-field equations. 
The physical solution is the one with the lowest ground state
energy.  To find it, we must evaluate the energy in the mean field
approximation.  There are different ways to address this.  Here we show how to calculate the ground state energy ``directly'', i.e. by taking the expectation value of the Hamiltonian.  We arrive at Eqs.~\eqref{eq:energy-av} and \eqref{eq:energy-kin} given in the main text.

We write the Hamiltonian in Eq.~\eqref{eq:ham-spinons} as
\begin{equation}
  \label{eq:47}
  H = \sum_{\mathbf{r} \in {\rm I},{\rm II}} \frac{J_{zz}}{2} Q_\mathbf{r}^2 + \mathcal{H}[\Phi,{\sf s}].
\end{equation}
We can treat the gMFT as a variational calculation, with the trial
wavefunction being the ground state of the gMFT Hamiltonian.  The
expectation value of the second term is straightforward, but that of the
first is not.  We need to write the Hamiltonian generalization of the
approximation of ``softening'' the $|\Phi|=1$ constraint in the path
integral.  

To do this, we write $\Phi_\mathbf{r} = x_\mathbf{r} + i y_\mathbf{r}$, and introduce canonical
momenta $p_{x,\mathbf{r}}, p_{y,\mathbf{r}}$, such that $[x_\mathbf{r}, p_{x,\mathbf{r}}]=i$ etc.  Then if
we write $\Pi_\mathbf{r} = p_{x,\mathbf{r}}+i p_{y,\mathbf{r}}$, the equivalent approximation in
Hamiltonian form is
\begin{equation}
  \label{eq:48}
  H \rightarrow \sum_{\mathbf{r} \in {\rm I},{\rm II}} \left\{ \frac{J_{zz}}{2} \Pi_\mathbf{r}^\dagger \Pi_\mathbf{r}^{\vphantom\dagger} +
  \lambda (\Phi_\mathbf{r}^\dagger \Phi_\mathbf{r}^{\vphantom\dagger}-1) \right\} +
  \mathcal{H}[\Phi,{\sf s}],
\end{equation}
since 
\begin{equation}
\Pi^\dagger\Pi\sim-\frac{1}{|\Phi|}\frac{\partial}{\partial|\Phi|}\left(|\Phi|\frac{\partial}{\partial|\Phi|}\right)+\frac{1}{|\Phi|^2}Q^2\sim Q^2,
\end{equation} 
because $|\Phi|=1$, as enforced by $\lambda (\Phi_\mathbf{r}^\dagger \Phi_\mathbf{r}^{\vphantom\dagger}-1)$ with $\lambda>0$. One can check that, writing the path integral for the Hamiltonian Eq.~\eqref{eq:48} and
integrating out $\Pi_\mathbf{r}$, one obtains the action used
previously.  Note that we can see that there should be four harmonic
oscillators per unit cell, arising from
$p_{x,\mathbf{r}},p_{y,\mathbf{r}}$ for the two sites $\mathbf{r}$ in
the diamond basis.  This implies in turn that each of  the two spinon branches
$\omega^\pm_\mathbf{k}$ is doubly degenerate.  

Returning to the energy calculation,  we simply can take the expectation value of Eq.~\eqref{eq:48}.
Note that, because we choose $\langle \Phi_\mathbf{r}^\dagger\Phi_\mathbf{r}\rangle = 1$,
the expectation value of the second term in the brackets (with the
$\lambda$ coefficient) is zero.  Hence the energy is
\begin{equation}
  \label{eq:49}
  E_{\rm GS} = \sum_{\mathbf{r} \in {\rm I},{\rm II}}  \frac{J_{zz}}{2} \left\langle
      \Pi_\mathbf{r}^\dagger \Pi_\mathbf{r}^{\vphantom\dagger} \right\rangle +
    \left\langle\mathcal{H}[\Phi,{\sf s}] \right\rangle = E_{kin}+ E_{av}.
\end{equation}
The second term is
\begin{eqnarray}
  \label{eq:50}
  E_{av} & = & \left\langle\mathcal{H}[\Phi,{\sf s}] \right\rangle \\
  & = & N_{u.c.}\left[ - 2 I_2 \cos^2\theta J_\pm - 4 I_1
    \sin2\theta J_{z\pm}\right], \nonumber
\end{eqnarray}
i.e. half of the constant term obtained from the replacement Eq.~\eqref{eq:MFTreplace} (up to a minus sign).
The ``kinetic energy'' $E_{kin}$ requires more thought.  We can calculate
it by the path integral.  One obtains
\begin{equation}
  \label{eq:51}
  \left\langle
      \Pi_\mathbf{r}^\dagger \Pi_\mathbf{r}^{\vphantom\dagger} \right\rangle =
    \frac{1}{J_{zz}}\left[ 2\delta(\tau=0) - \frac{\langle \partial_\tau
        \Phi^\dagger_\mathbf{r} \partial_\tau \Phi_\mathbf{r}^{\vphantom\dagger}\rangle}{J_{zz}}\right].
\end{equation}
Each of the two terms in the square brackets is formally divergent, but
together they give a finite answer.  This can be seen by writing it all
in Fourier space:
\begin{eqnarray}
  \label{eq:52}
  E_{kin} & = & \frac{J_{zz}}{2} N_{u.c.} \sum_{\mathbf{r}\in{\rm I},{\rm II}} \left\langle
    \Pi_\mathbf{r}^\dagger \Pi_\mathbf{r}^{\vphantom\dagger} \right\rangle \\
  & = & \frac{2N_{u.c.}}{2J_{zz}} \int \! \frac{d\omega_n}{2\pi} \int_\mathbf{k} \left[
    2J_{zz} - \omega_n^2 \left[G_{\mathbf{k},\omega_n}\right]_{11}\right]. \nonumber 
\end{eqnarray}
Now we can rewrite the Green's function as
\begin{eqnarray}
  \label{eq:53}
  \left[G_{\mathbf{k},\omega_n}\right]_{11} & = & \frac{\Omega_n^2 + \frac{1}{2}(z_+^2 +
    z_-^2)}{(\Omega_n^2 + z_+^2)(\Omega_n^2+z_-^2)},
\end{eqnarray}
where $\Omega_n = \omega_n/\sqrt{2J_{zz}}$, and $z_\pm=z^\pm_\mathbf{k}$.  Changing variables from
$\omega_n$ to $\Omega_n$, we have 
\begin{eqnarray}
  \label{eq:54}
  E_{kin} & = & 2 N_{u.c.} \sqrt{2J_{zz}}\int_\mathbf{k} \int \! \frac{d\Omega_n}{2\pi} 
  \frac{\Omega_n^2 (z_+^2+z_-^2)/2 + z_+^2 z_-^2}{(\Omega_n^2 +
    z_+^2)(\Omega_n^2+z_-^2)} \nonumber \\
  &=& \frac{N_{u.c.}}{2} \int_\mathbf{k} [\omega^+_\mathbf{k} + \omega^-_\mathbf{k}].
\end{eqnarray}

We note that this formulation makes it clear that this form of the energy is variational.
Specifically, even if $\theta$ is {\sl not} chosen to satisfy the mean
field condition, this should give an upper bound to the ground state
energy.  This means that this form of the energy should be minimized
when $\theta$ equals its true value, which is the minimum energy mean
field solution.  So this form could be used, if desired, to {\sl search}
for the ground state by just minimizing this energy as a function of
$\theta$.  Note that we do need to choose $\lambda$ as a function of
$\theta$ to enforce the $|\Phi|=1$ condition for this to be valid, though.

\section{Phase Boundaries}

Most generally, the phase transitions of Figure \ref{fig:MF-phasediag} are found numerically as the loci of points where the lowest-energy solutions to the gMFT equations have different $(\theta,\lambda)$ characteristics on either side of the transition line.  It turns out that some of these lines coincide with curves which can be found more directly.  Here we outline which transitions have a direct physical meaning, and how we find them.

We define, and shall use extensively, the dimensionless exchange constants
\begin{equation}
\label{eq:dimensionless}
\tilde{J}_i=\frac{J_i}{J_{zz}}.
\end{equation}
Finally, the problem being symmetric in $J_{z\pm}\rightarrow-J_{z\pm}$, we will always assume $J_{z\pm}\geq0$.  In the latter case, all the values of $\theta$ of interest are contained in the $[0,\pi/2[$ interval, so that $J_2$, defined in Eq.~\eqref{eq:beta}, is always positive.  Hence, $|J_2|=J_2$.

\subsection{QSL - AFM}

In both the QSL and AFM, $\theta=0$.  This considerably simplifies the problem.  Using Eqs.~\eqref{eq:consistency} and \eqref{eq:integrals}, we find that the maximum allowed $J_{\pm}$ {\sl within the QSL} is such that $\lambda=3J_{\pm}$, from which
\begin{equation}
\label{eq:QSL-AFM}
\left(\frac{J_{\pm}}{J_{zz}}\right)_{{\rm max}}^{\rm QSL}=\frac{1}{2}\left(\int_\mathbf{k}\frac{1}{\sqrt{3-\frac{1}{2}L_\mathbf{k}}}\right)^2
\end{equation}
follows.  The actual QSL-AFM transition happens when $\frac{J_{\pm}}{J_{zz}}=\left(\frac{J_{\pm}}{J_{zz}}\right)_{{\rm max}}^{\rm QSL}$.

\subsection{QSL - CFM}

We find that the QSL-CFM transition is continuous in $\theta$.  We can thus find the transition line as follows.  We expand the consistency equations Eqs.~\eqref{eq:consistency} to first order in $\theta$, around $\theta=0$,
\begin{eqnarray}
I_1(\theta,\lambda)&=&\theta \left|J_{z\pm}\right|\hat{I}_1(\lambda)+O(\theta^2)\\
I_2(\theta,\lambda)&=&I_2(0,\lambda)+O(\theta^2)\\
I_3(\theta,\lambda)&=&I_3(0,\lambda)+O(\theta^2)
\end{eqnarray}
where $\hat{I}_1(\lambda)=\frac{1}{2}\sqrt{\frac{J_{zz}}{2}}\left(\frac{Z_\mathbf{k}^+(\theta=0,\lambda)}{2}\right)^{3}$ (note that $\hat{I}_1$ has the dimension of $1/J$), so that Eqs.~\eqref{eq:consistency} become
\begin{equation}
\begin{cases}
J_\pm I_2(0,\lambda)=2J_{z\pm}^2\hat{I}_1(\lambda)\\
I_3(0,\lambda)=1
\end{cases},
\end{equation}
where neither $I_2(0,\lambda)$, $I_3(0,\lambda)$ nor $\hat{I}_1(\lambda)$ depend on $J_{z\pm}$.  For every value of $J_\pm$, $\lambda$ is the solution of $I_3(0,\lambda)=1$, so that we find $J_{z\pm}/J_{zz}$ at the transition to be
\begin{equation}
\left(\frac{J_{z\pm}}{J_{zz}}\right)_{crit}^{\rm QSL-CFM}=\sqrt{\frac{\tilde{J}_\pm I_2(0,\lambda)}{2J_{zz}\hat{I}_1(\lambda)}},
\end{equation}
with $\lambda(J_\pm)$ first determined by solving $I_3(0,\lambda)=1$ at fixed $J_\pm$.

\subsection{CFM - FM}

The CFM-FM transition coincides with the first appearance of a condensed solution (at $J_\pm$ fixed for increasing $J_{z\pm}$) to the gMFT equations Eqs.~\eqref{eq:consistency}.  To find this transition line, we use the generic form of $\ell_\mathbf{k}^\pm=J_1 L_\mathbf{k}\mp|J_2 M_\mathbf{k}|$.  In other words, we treat $J_1$ and $J_2$ defined in Eqs.~\eqref{eq:alpha} and \eqref{eq:beta} as parameters.  We also define $r=J_2/J_1=2\frac{J_{z\pm}}{J_\pm}\tan\theta$, so that
\begin{equation}
\ell^\pm_\mathbf{k}=J_1 L_\mathbf{k}\mp|J_2 M_\mathbf{k}|=J_1\left(L_\mathbf{k}\mp r |M_\mathbf{k}|\right).
\end{equation}

The system being at the onset of a {\sl possible} condensate, we know the possible values of $\lambda$ at this onset (see above section, Eq.~\eqref{eq:lambdamin}), 
\begin{equation}
\begin{cases}
\lambda=2J_1\left(2r-1\right)\\
\mbox{or}\;\lambda=J_1\left(6+\frac{r^2}{2}\right)
\end{cases},
\end{equation}
so that, from $I_3(\theta,\lambda)=1$ (with these values of $\lambda$), we get either
\begin{equation}
\label{eq:alpha-r1}
J_1\left(r\right)=\frac{J_{zz}}{8}\left[\int d^3k\sum_{\nu=\pm1}\frac{1}{\sqrt{r(4+\nu |M_\mathbf{k}|)-2-L_\mathbf{k}}}\right]^2,
\end{equation}
or
\begin{equation}
\label{eq:alpha-r2}
J_1\left(r\right)=\frac{J_{zz}}{8}\left[\int d^3k\sum_{\nu=\pm1}\frac{1}{\sqrt{6-L_\mathbf{k}+r\left(\frac{1}{2}r+\nu |M_\mathbf{k}|\right)}}\right]^2,
\end{equation}
i.e. $J_1$ as a function of $r$.  Using the above equations, we can
rewrite the consistency equation Eq.~\eqref{eq:consistency} in a form
suitable for determining $r$ as an unknown. For example, in the first
case Eq.~\eqref{eq:alpha-r1}, the transition line is best found as a
function of $J_\pm$, so we fix $J_\pm$, and use the simple relations
(obtained from Eqs.~\eqref{eq:alpha} and \eqref{eq:beta})
\begin{eqnarray}
\theta&=&\arccos\sqrt{\frac{2J_1}{J_\pm}}\\
J_{z\pm}&=&\frac{1}{2}r\frac{1}{\tan\theta}J_\pm.
\end{eqnarray}
Using these relations in Eq.~\eqref{eq:consistency} and
Eq.~\eqref{eq:alpha-r1}, we obtain an equation which can be solved
numerically for $r$ as a function of $J_\pm$. Then, by substituting
back, we get $J_{z\pm}$ and hence the phase boundary in the
$J_{\pm}$--$J_{z\pm}$ plane.  In the second case Eq.~\eqref{eq:alpha-r2},
we fix $\theta$ (as opposed to $J_\pm$), use the relations
\begin{eqnarray}
J_\pm&=&\frac{2J_1}{\cos^2\theta}\\
J_{z\pm}&=&\frac{1}{2}r\frac{1}{\tan\theta}J_\pm,
\end{eqnarray}
and proceed similarly.

\subsection{CFM - AFM and AFM - FM}

The transitions are here discontinuous in $\theta$, and we find the transitions indirectly, except for a small portion of the CFM-AFM transition line, which coincides with the limiting boundary described above, with $\lambda=6J_1+\frac{J_2^2}{2J_1}$.

\section{Benchmarking gMFT}

\subsection{Classical approach}

\begin{figure*}[htbp]
\begin{center}
\includegraphics[width=.8\textwidth]{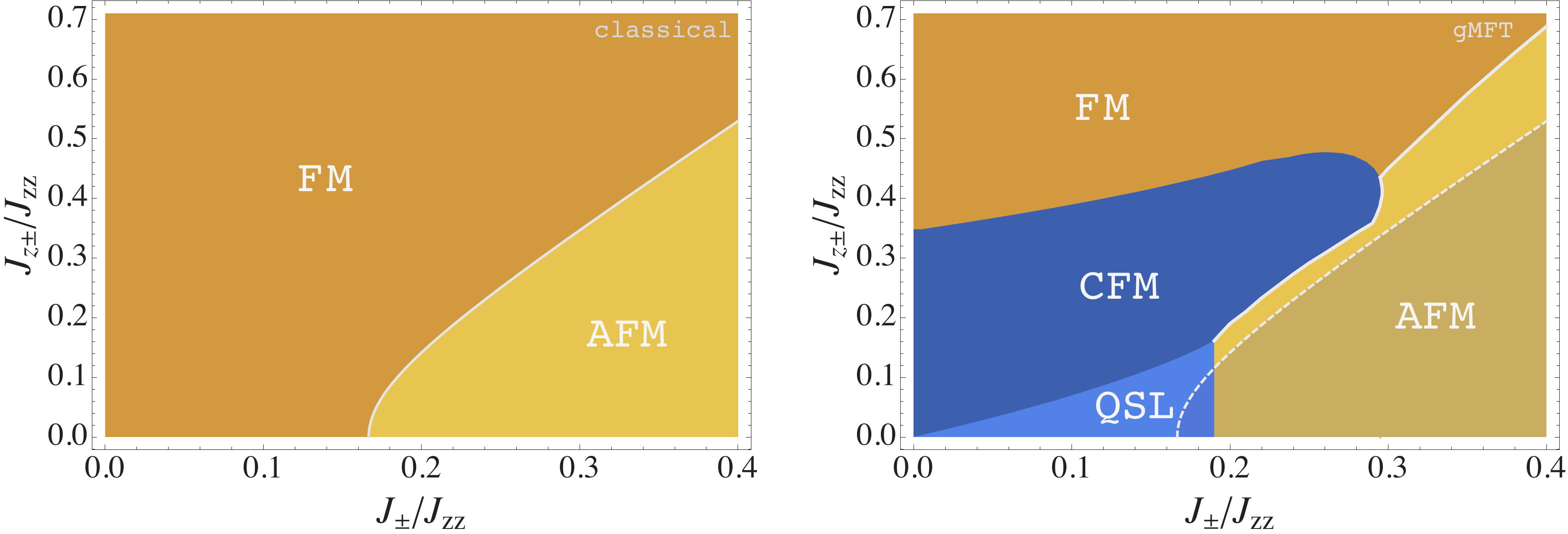}
\caption{$T=0$ classical phase diagram (left), and classical phase diagram superimposed over that of gMFT (right).}
\label{fig:classical-phasediag}
\end{center}
\end{figure*}

In Figure \ref{fig:classical-phasediag}, we compare the phase diagram of the gMFT with that expected by treating the spins in Eq.~\eqref{eq:spinham} classically.  The large $J_\pm/J_{z\pm}$ region is found to be antiferromagnetic, while the large $J_{z\pm}/J_{\pm}$ is ferromagnetic.  

The transition line is found analytically.  To achieve this, we use the fact that, both in the FM and AFM phases, the classical spins take a specific and simple form.

In the FM phase,
\begin{equation}
\begin{cases}
\vec{S}_0^{\rm FM}=(x\mathbf{\hat{a}}_0+y\mathbf{\hat{b}}_0+z\mathbf{\hat{e}}_0)/2\\
\vec{S}_1^{\rm FM}=(-x\mathbf{\hat{a}}_1-y\mathbf{\hat{b}}_1-z\mathbf{\hat{e}}_1)/2\\
\vec{S}_2^{\rm FM}=(-x\mathbf{\hat{a}}_2-y\mathbf{\hat{b}}_2-z\mathbf{\hat{e}}_2)/2\\
\vec{S}_3^{\rm FM}=(x\mathbf{\hat{a}}_3+y\mathbf{\hat{b}}_3+z\mathbf{\hat{e}}_3)/2
\end{cases},
\end{equation} 
where $x,y,z$ are free parameters with $x^2+y^2+z^2 = 1$.  In polar
coordinates, we write $x=\sin\Theta\cos\phi, y = \sin\Theta\sin\phi,
z=\cos\Theta$, and the energy Eq.~\eqref{eq:spinham} per unit cell becomes
\begin{eqnarray}
\epsilon^{\rm FM}_{class}&=&\frac{1}{8}\Big[-2\cos^2\Theta J_{zz}+4\sin^2\Theta J_\pm\\
&&\qquad\left.+4\sin2\Theta\left(\cos\phi+\sqrt{3}\sin\phi\right)J_{z\pm}\right].\nonumber
\end{eqnarray}
The energy is then found to be minimized for $\phi=\pi/3$ and $\Theta=\arctan\left[\frac{1+2\tilde{J}_\pm-\sqrt{1+4\tilde{J}_\pm(\tilde{J}_\pm+1)+64\tilde{J}_{z\pm}^2}}{8\tilde{J}_{z\pm}}\right]$, which yields $\epsilon^{\rm FM}_{class}=\dfrac{J_{zz}}{8}\left[-1+2\tilde{J}_\pm-\sqrt{(1+2\tilde{J}_\pm)^2+64\tilde{J}_{z\pm}^2}\right]$, where $\tilde{J}_i=J_i/J_{zz}$, Eq.~\eqref{eq:dimensionless}.

In the AFM phase, the solution takes the form
\begin{equation}
\vec{S}_\mu^{\rm FM}=\frac{1}{2}\left(X\mathbf{\hat{a}}_\mu+Y\mathbf{\hat{b}}_\mu\right),
\end{equation}
with arbitrary $X,Y$ such that $X^2+Y^2=1$.  This gives the energy per
unit cell
\begin{equation}
\epsilon^{\rm AFM}_{class}=-\frac{3J_\pm}{2}.
\end{equation}
Equating $\epsilon^{\rm FM}_{class}=\epsilon^{\rm AFM}_{class}$, we readily find that the transition line takes the form
\begin{equation}
\left(\frac{J_{z\pm}}{J_{zz}}\right)_{crit}^{class}=\sqrt{\frac{\tilde{J}_\pm(6\tilde{J}_\pm-1)}{2}}.
\end{equation}
The comparison between the classical and the gMFT diagrams shows that the two approaches agree in the semiclassical limit of the gMFT, i.e. for $J_\pm/J_{zz},J_{z\pm}/J_{zz}\gg1$.


\subsection{Perturbative approach}

\begin{figure}[htbp]
\begin{center}
\includegraphics[width=3.3in]{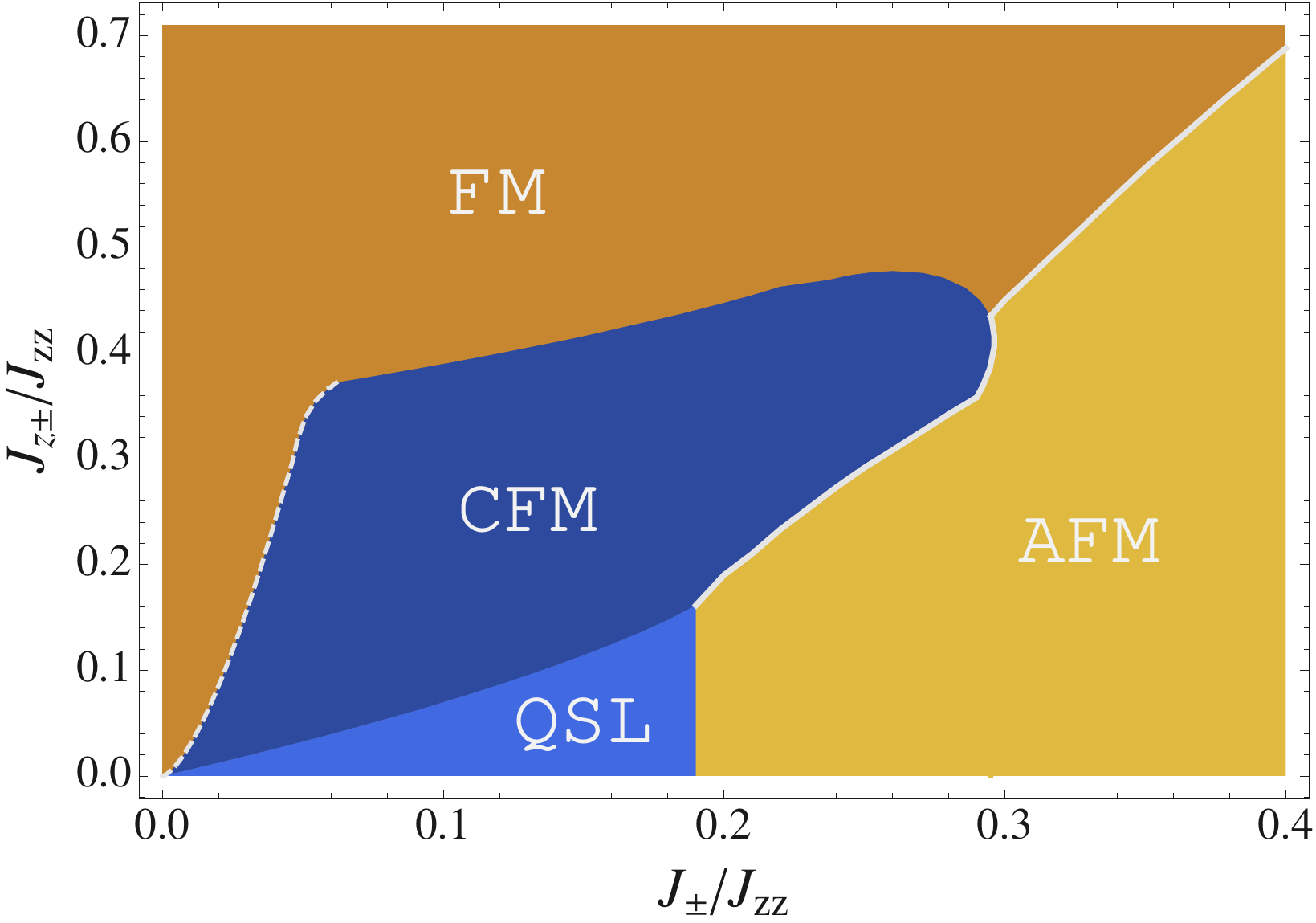}
\caption{Modified gMFT diagram, which takes into account the known perturbative limit $J_{z\pm}/J_{zz}\ll1$. Note that the FM-CFM transition (white dashed line) in the latter region is a {\sl sketch}.}
\label{fig:modified-gMFT-diag}
\end{center}
\end{figure}

In the small $J_{z\pm}/J_{zz}$ and $J_{\pm}/J_{zz}$ limit, the perturbative approach developed in Refs.~\onlinecite{hermele} and \onlinecite{jointpaper} applies.  There, the effective Hamiltonian is
\begin{eqnarray}
H_{\rm eff}^{\rm pert}&=&-K\sum_{\{i,j,k,l,m,n\}=\hexagon}\left(\mathsf{S}_i^+\mathsf{S}_j^-\mathsf{S}_k^+\mathsf{S}_l^-\mathsf{S}_m^+\mathsf{S}_n^-+\mbox{h.c.}\right)\nonumber \\
&& \;-J_{(3)}\sum_{\langle\langle\langle i,j\rangle\rangle\rangle}\mathsf{S}_i^z\mathsf{S}_j^z,
\end{eqnarray}
with $K=\dfrac{12J_\pm^3}{J_{zz}^2}$ and $J_{(3)}=\dfrac{3J_{z\pm}^2}{J_{zz}}$, and $\hexagon$ represents the (flat) hexagons of the {\sl pyrochlore} lattice.  While the first term induces the spin liquid physics, the second, taken alone has the six ferromagnetic ground states described in Ref.~\onlinecite{jointpaper}, with polarization along the $\langle100\rangle$ directions.  The actual phase diagram is therefore expected to differ from that of the gMFT in the small $J_{z\pm}/J_{zz}$ region (but agree when $J_{z\pm}/J_{zz}=0$).  Dimensionally we find that the transition between the FM and the CFM in this region occurs along
\begin{equation}
\left(\frac{J_{z\pm}}{J_{zz}}\right)_{crit}^{\rm PT}\sim\left(\frac{J_{\pm}}{J_{zz}}\right)^{3/2},
\end{equation}
in the region where perturbation theory applies. A {\sl sketch} of the expected gMFT diagram which includes this perturbative limit is sketched in Figure \ref{fig:modified-gMFT-diag}.

\section{Physical Properties}

\subsection{Staggered magnetization}

Here we calculate the ``staggered'' magnetization within the QSL and AFM phases, i.e. the {\sl local} magnetization of each spin. This quantity will allow us to distinguish between the QSL (uncondensed) and AFM (condensed) since the reasons for overall zero magnetization in both phases are different. Indeed, in the AFM, the local the spin expectation values are non-zero but compensate, while in the QSL, the (local) XY symmetry remains unbroken, and $\langle \vec{\mathsf{S}}_i\rangle=\vec{0}$ everywhere. 

\begin{figure}[htbp]
\begin{center}
\includegraphics[width=3.3in]{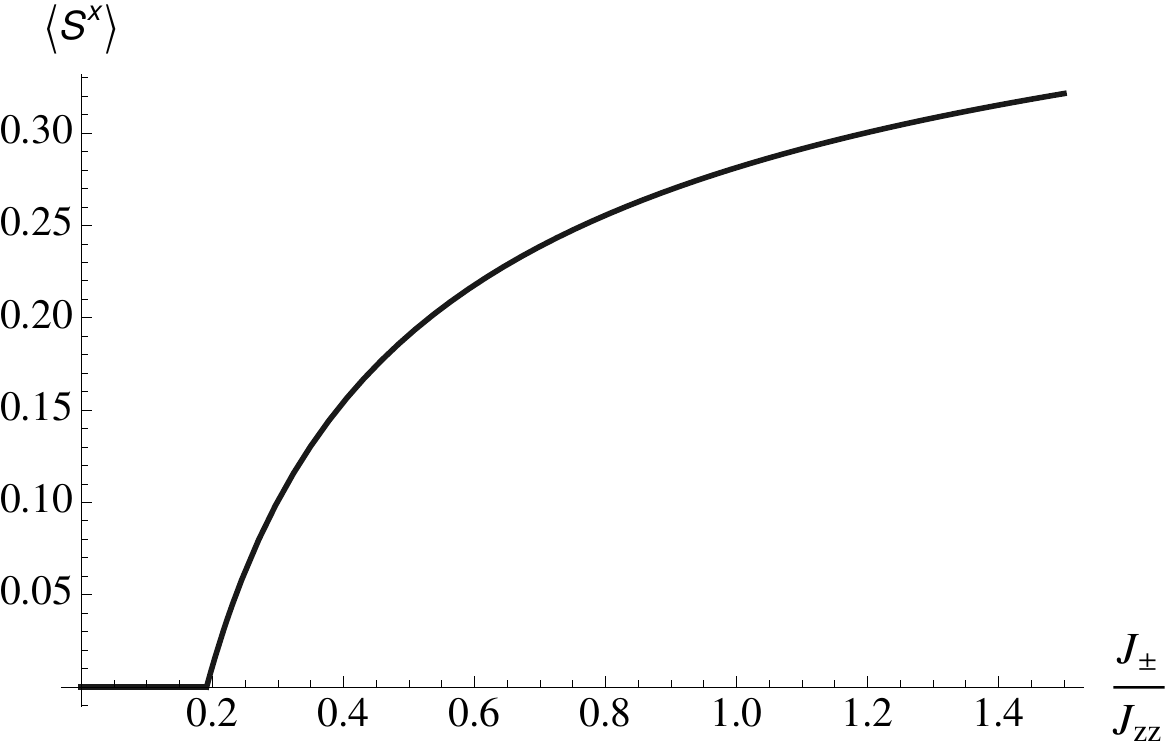}
\caption{Staggered magnetization $\langle\mathsf{S}^x\rangle$ as a function of $J_\pm/J_{zz}$, Eq.~\eqref{eq:stagmag}, for $J_{z\pm}=0$.}
\label{fig:staggered-mag}
\end{center}
\end{figure}

We calculate the staggered magnetization along the local $x$ axes
.  To do so, we compute the correlation function $\langle\mathsf{S}_i^+\mathsf{S}_j^-\rangle$, where sites $i$ and $j$ are taken far apart from one another, where they can be considered independent.
\begin{eqnarray}
\langle\mathsf{S}_i^+\mathsf{S}_j^-\rangle&=&\langle\Phi_{\mathbf{r}}^\dagger\,\mathsf{s}^+_{\mathbf{r},\mathbf{r}+\mathbf{e}_\mu}\Phi_{\mathbf{r}+\mathbf{e}_\mu}\Phi_{\mathbf{r}'+\mathbf{e}_\nu}^\dagger\mathsf{s}^-_{\mathbf{r}',\mathbf{r}'+\mathbf{e}_\nu}\Phi_{\mathbf{r}'}\rangle\\
&=&\langle\mathsf{s}_\mu^+\rangle\langle\mathsf{s}_\nu^-\rangle\langle\Phi_{\mathbf{r}}^\dagger\Phi_{\mathbf{r}'}\rangle\langle\Phi_{\mathbf{r}+\mathbf{e}_\mu}\Phi_{\mathbf{r}'+\mathbf{e}_\nu}^\dagger\rangle,
\end{eqnarray}
where $\mathbf{r}$ and $\mathbf{r}'$ are on diamond sublattice I and, of course, $i\in\langle\mathbf{r},\mathbf{r}+\mathbf{e}_\mu\rangle$ and $j\in\langle\mathbf{r}',\mathbf{r}'+\mathbf{e}_\nu\rangle$, and where we made use of Wick's theorem. From Eq.~\eqref{eq:green-fn}, we find
\begin{eqnarray}
\langle\Phi_{\mathbf{r}}^\dagger\Phi_{\mathbf{r}'}\rangle&=&\frac{1}{2N_{u.c.}}\sqrt{\frac{J_{zz}}{2}}\sum_\mathbf{k}Z_{\mathbf{k}}^+(0,\lambda)e^{i\mathbf{k}\cdot(\mathbf{r}'-\mathbf{r})}.
\end{eqnarray}
In the QSL, the sum can be replaced by its corresponding integral,
\begin{equation}
\frac{1}{N_{u.c.}}\sum_\mathbf{k}Z_{\mathbf{k}}^+(0,\lambda)e^{i\mathbf{k}\cdot(\mathbf{r}'-\mathbf{r})}\rightarrow\int_\mathbf{k}Z_{\mathbf{k}}^+(0,\lambda)e^{i\mathbf{k}\cdot(\mathbf{r}'-\mathbf{r})}=0,
\end{equation}
because $|\mathbf{r}'-\mathbf{r}|\rightarrow\infty$, and $z^-_\mathbf{k}$ never reaches $0$ within the integration domain. We thereby recover $\langle\mathsf{S}^x\rangle=0$ in the QSL. In the AFM, however, the ``minimum term'' survives via the subextensive part of $\lambda$, Eq.~\eqref{eq:59}, so that 
\begin{eqnarray}
\langle\Phi_{\mathbf{r}}^\dagger\Phi_{\mathbf{r}'}\rangle&=&\sqrt{\frac{J_{zz}}{2}}\frac{1}{\delta},
\end{eqnarray}
since $\mathbf{k}_{\rm min}=\mathbf{0}$ when $\theta=0$.  Using Eq.~\eqref{eq:60}, we find
\begin{eqnarray}
\sqrt{\frac{J_{zz}}{2}}\frac{1}{\delta}&=&1-\sqrt{\frac{J_{zz}}{2}}\left[\int_\mathbf{k}\frac{1}{\sqrt{\lambda_{\rm min}-\frac{1}{2}J_\pm L_\mathbf{k}}}\right]\\
&=&1-\sqrt{\frac{J_{zz}}{J_\pm}}\sqrt{\left(\frac{J_\pm}{J_{zz}}\right)_{\rm max}^{\rm QSL}},
\end{eqnarray}
as defined in Eq.~\eqref{eq:QSL-AFM}.  Similarly, $\langle\Phi_{\mathbf{r}+\mathbf{e}_\mu}\Phi_{\mathbf{r}'+\mathbf{e}_\nu}^\dagger\rangle=\sqrt{\frac{J_{zz}}{2}}\frac{1}{\delta}$ in the AFM, so that
\begin{equation}
\label{eq:stagmag}
\langle\mathsf{S}^x\rangle=\sqrt{\langle\mathsf{S}_i^+\mathsf{S}_j^-\rangle}=\frac{1}{2}\left(1-\sqrt{\frac{\tilde{J}_\pm^c}{\tilde{J}_\pm}}\right),
\end{equation}
in the AFM phase for $J_{z\pm}=0$, where we have defined $\tilde{J}_\pm=J_\pm/J_{zz}$ and $\tilde{J}_\pm^c=\left(J_\pm/J_{zz}\right)_{\rm max}^{\rm QSL}$.  $\langle\mathsf{S}^x\rangle$ at $J_{z\pm}=0$ is plotted versus $J_\pm/J_{zz}$ in Figure \ref{fig:staggered-mag}.

\subsection{Structure factor}

The excitations above the uncondensed ground states of
Eq.~\eqref{eq:ham-spinons} comprise the photon, spinons and ``electric''
monopoles \cite{hermele,jointpaper}.  In general, the spinons and
monopoles are gapped excitations, and so do not contribute to the low-energy part of the structure factor.   Below some threshold, only the
photon survives, and gives rise to a universal form of the scattering.  Here we
sketch the general argument which shows that it appears in the inelastic
structure factor as a linearly dispersing mode at arbitrarily low
energy, with an unusually small spectral weight and polarization
dependence which distinguishes it from the usual spin waves.  We will be
deliberately vague here about lattice details, orientation of local spin
axes, etc, as these affect only $O(1)$ dependencies of the amplitudes
and not the important scaling properties.  A much more complete and
microscopic study of the structure factor in the QSL and CFM phases is
planned for a future publication.

Rather than derive it microscopically, we rely on the universality of
the low energy description of the emergent photon.  We expect that it is
described by the usual electromagnetic action, which may be written in
the continuum for small momentum and low energy. We write this in
Euclidean time in terms of the vector potential, and choose the gauge in
which the scalar potential vanishes.  The result is familiar:
\begin{equation}
  \label{eq:2}
  \mathcal{S}_{\rm QED} = \int_{\tau {\bf r}} \Big[ \frac{c_1}{2} (\partial_\tau
  {\bf A})^2 + \frac{c_2}{2} ({\boldsymbol \nabla}\times {\bf A})^2 \Big].
\end{equation}
Here $c_1$ and $c_2$ are phenomenological parameters related to the
effective dielectric constant and magnetic permeability for the
effective electrodynamics.  Importantly, the photon velocity is given by
$v=\sqrt{c_2/c_1}$.  We recall the relation between the formalism of
Eq.~\eqref{eq:ham-spinons} and that of lattice electrodynamics, given in
the main text,
\begin{equation}
\begin{cases}
\mathsf{S}^z_{\mathbf{rr}'}=\mathsf{s}^z_{\mathbf{rr}'}=E_{\mathbf{rr}'}\\
\mathsf{S}^\pm_{\mathbf{rr}'}=\Phi_\mathbf{r}^\dagger\,\mathsf{s}^\pm_{\mathbf{rr}'}\Phi_{\mathbf{r}'}= \Phi_\mathbf{r}^\dagger\,e^{\pm iA_{\mathbf{rr}'}}\Phi_{\mathbf{r}'}
\end{cases},
\label{eq:spins-ELD}
\end{equation}
where $\mathbf{r}$ is on diamond sublattice I.  Because the transverse
components of the spins, $\mathsf{S}^\pm$, involve the spinon fields, we
expect that they contribute only above the two-spinon threshold.  For
low energies, it suffices to consider $\mathsf{S}^z$ only.  We take the
continuum limit by defining the electric field to be ``oriented'' along
the diamond lattice bond $\mbox{I}\rightarrow\mbox{II}$ on which the
spin lies and centered midway along it, i.e.
\begin{equation}
{\bf E}_{\mathbf{r}+\mathbf{e}_\mu/2}=E_{\mathbf{r},\mathbf{r}+\mathbf{e}_\mu}\mathbf{\hat{e}}_\mu,
\end{equation}
Similarly, recall that the magnetic moment due to
$\mathsf{S}^z_{\mathbf{r},\mathbf{r}+\mathbf{e}_\mu}$ is along the
$\mathbf{\hat{e}}_\mu$ direction in real space,
\begin{equation}
  \label{eq:3}
  \mathbf{S}_{\mathbf{r}+{\mathbf{e}_\mu}/2} = {\mathsf
    S}^z_{\mathbf{r},\mathbf{r}+\mathbf{e}_\mu}  {\bf\hat e}_\mu = {\bf
    E}_{\mathbf{r}+\mathbf{e}_\mu/2} .
\end{equation}
So we see that, in this low energy subspace where excited spinons
may be neglected, the spin operator is precisely the electric field
operator.  Then, using the usual relation
$\mathbf{E}(\mathbf{r})=-\boldsymbol{\nabla}_\mathbf{r}V(\mathbf{r})-\partial_\tau
\mathbf{A}(\mathbf{r})$ for $V=0$ (gauge choice), 
\begin{equation}
  \label{eq:4}
   {\mathbf S}_{\mathbf{k},\omega_n} = i\omega_n {\mathbf A}_{\mathbf{k},\omega_n},
\end{equation}
where $\omega_n$ is the bosonic Matsubara frequency. Thus the Matsubara correlation function of the spins is
\begin{equation}
\label{eq:6}
\langle S^i_{-\mathbf{k},-\omega_n}S^j_{\mathbf{k},\omega_n}\rangle=\omega_n^2
\langle
A^i_{-\mathbf{k},-\omega_n} A^j_{\mathbf{k},\omega_n}\rangle,
\end{equation}
Calculation of the gauge field propagator is a textbook exercise:
\begin{equation}
  \label{eq:5}
  \left\langle A^i_{-\mathbf{k}-\omega} A^j_{\mathbf{k}\omega}\right\rangle =
    c_1^{-1}\frac{\omega_n^2 \delta_{ij}+v^2 k_i k_j }{\omega_n^2 + v^2
      \mathbf{k}^2}.
\end{equation}
Inserting this into Eq.~\eqref{eq:6} and analytically continuing
$i\omega_n \rightarrow \omega+i0^+$, we arrive finally at the inelastic
structure factor
\begin{eqnarray}
  \label{eq:7}
{\mathcal F}_{\mathbf{k},\omega} &=& -\left.{\rm Im}\left[
   \langle S^i_{-\mathbf{k},-\omega_n}S^j_{\mathbf{k},\omega_n}\rangle\right]\right|_{i\omega_n
 \rightarrow \omega+i0^+}
  \\
&\sim& \frac{\pi }{c_1} \big[ \delta_{ij} -
  \frac{k_i k_j}{\mathbf{k}^2}\big] \; \omega \,\delta( \omega-v |\mathbf{k}|).\nonumber
\end{eqnarray}
We see that the photon appears as a sharp peak in the structure factor,
with a weight proportional to its frequency $\omega$.  This indicates,
as mentioned in the text, a strong suppression of the weight at low
energy, especially relative to the familiar case of a spin wave, for
which the weight {\sl diverges} like $1/\omega$ in the same limit.


\end{document}